\documentclass[aps,prd,nofootinbib,floatfix,twocolumn]{revtex4}
\usepackage{amsmath,amssymb,color,graphicx,bm}
\usepackage{physics}

\definecolor{red}{rgb}{0.8,0,0}
\definecolor{RED}{rgb}{0.8,0,0}
\definecolor{violet}{rgb}{0.4,0,0.4}
\definecolor{green}{rgb}{0,0.5,0.0}
\definecolor{GREEN}{rgb}{0,0.5,0.0}
\definecolor{navy}{rgb}{0.0,0.0,0.6}
\definecolor{orange}{rgb}{0.8,0.2,0.0}
\definecolor{blue}{rgb}{0.3,0.0,0.8}

\begin{document}

\title{\textbf{Low frequency gravitational waves emerge Berry phase}}
\author{Partha Nandi$^{a,b}$}
\email{parthanandyphysics@gmail.com}
\author{Sounak Pal$^c$}
\email{pal.rik99@gmail.com}
\author{Sayan Kumar Pal$^{a,d}$}
\email{pal.sayan566@gmail.com}
\author{Bibhas Ranjan Majhi$^e$}
\email{bibhas.majhi@iitg.ac.in}
\affiliation{$^a$S N Bose National Centre or Basic Sciences,
Block-JD, Sector-III, Saltlake - 700106, India\\
$^b$ Institute of Theoretical Physics, University of Stellenbosch, Stellenbosch-7600, South Africa\\
$^c$Department of Physics,
Ramakrishna Mission Vivekananda Educational and Research Institute
Belur Math, Howrah-711202, India\\
$^d$Physics and Applied Mathematics Unit, Indian Statistical Institute, Kolkata,
203 B.T. Road, West Bengal 700108, India\\
$^e$Department of Physics, Indian Institute of Technology Guwahati, Guwahati 781039, Assam, India.}

\begin{abstract}
	The detection of low frequency gravitational waves (LFGWs)  astronomy has marked an advent of new era in the domain of astrophysics and general relativity. Using the framework of interaction between GWs and a point two-particles like detector, within linearized gravity approach, we propose a toy detector model whose quantum state is being investigated at a low-frequency of GWs. The detector is in simultaneous interaction with GWs and an external time-dependent (tuneable) two-dimensional harmonic potential. We observe that the interaction with low frequency GWs naturally provides adiabatic approximation in the calculation, and thereby can lead to a quantal geometric phase in the quantum states of the detector. Moreover this can be controlled by tuning the frequency of the external harmonic potential trap. We argue that such geometric phase detection may serve as a manifestation of the footprint of  GWs. More importantly, our theoretical model may be capable of providing a layout for the detection of very small frequency GWs through Berry phase.
\end{abstract}
\maketitle	

\section{Introduction}
The ground based laser interferometric techniques employed in the LIGO and the Virgo
experiments have been phenomenally successful in the detection of gravitational waves (GWs) through a classical treatment of the arms of the interferometer \cite{gwdetect1}. The typical frequency range of detection of GWs in these experiments has been $5$ Hz -- $20,000$ Hz (e.g. see \cite{5Hz}). However, the European Space Agency launched the LISA Pathfinder mission in 2015 to test the technology required for a full-fledged space-based gravitational wave detector, with the goal of detecting much lower frequencies \cite{LK}.  In fact, it is anticipated that inflation in the early Universe is the source of primordial gravitational waves, which have a very low frequency \cite{LGWs}. It is crucial to find these gravitational waves in order to confirm the inflationary theory. In this article, we propose a theoretical model which has potential to be a candidate for experimental detection of such low frequency GWs (LFGWs) {\footnote{Onwards we will use LFGWs as the abbreviation for low frequency GWs.}}. 

 Usually GWs are detected through its interaction with lab apparatus like interferometer arms in LIGO. Particularly the very LFGWs are capable of providing adiabatic change in the detectors. A heuristic explanation is as follows. Consider GWs propagating along $z$-direction, whose form in linearized approximation in transverse-traceless (TT) gauge can be taken as $h_{ij}(t)\sim \cos(\omega_{g}t-kz)$. This induces a deviation of the trajectory of a point particle detector which is determined by $\dot{h}_{ij}(t)\sim\omega_{g} \sin(\omega_{g}t-kz)$ in the Hamiltonian (see, \cite{maggiore} for details). Therefore for the very low frequency range ($10^{-5}Hz <\omega_{g}<1$ Hz) the perturbations in Hamiltonian caused by the GWs are ultra slowly varying functions of time under adiabatic passage. This behavior can be quantified by a dimensionless parameter (we will denote this as $\epsilon$), defined through the system's internal and external time scales, which we will delve into in  details at the respective portion of our anlysis. Then these LFGWs are capable of inducing geometric phase (known as Berry phase (BP)) along with the usual dynamical phase in the quantum state of the detector. If this is true then the LFGWs can be distinguished through the BP.
 
Before describing our model to investigate the above geometric phase, let us now mention few earlier investigations on the gravitationally induced BP.  There has been an opinion among the physicists from quite sometime through various investigations that quantum mechanical domain will provide more prominent and experimentally tenable trace of gravitational waves on matter \cite{kieferweber}. It has also been known that gravitons exhibit Berry's geometric phase shift \cite{berry1} in the presence of a background Friedmann–Lemaitre–Robertson–Walker (FLRW) metric \cite{relicgravi}. Besides in \cite{relicgravi1}, a connection has been pointed out between the lower bound of von-Neumann entanglement entropy and the BP defined for quantum ground states of a generic solid state system. Such phase then serve as another plausible quantum fingerprint of the interaction of GWs with matter \cite{ecgs,wald2}.
 In fact, some investigations have been carried out how gravitational forces generally interact with quantum fluids. In \cite{meso2}, Anandan and Chiao investigated how by employing superfluids one can build antennas for gravitational radiation and then making use of superconducting circuits, it is possible to detect gravitational radiation \cite{meso3}. Interactions generated through the Lense-Thirring effect in rotating superconductors had been considered by De-Witt and Papini where they computed the resultant quantum phase shift \cite{dew,papi}. Apart from quantum fluids, classical Weber bar detectors have also been previously studied in the quantum regime by using quantum non-demolition measurements \cite{meso1}.

 Motivated by the above facts and investigations, we now propose the following theoretical model for a detector which changes adiabatically under the LFGWs and therefore is capable of acquire BP on its quantum state. The detector effectively consists of two uncoupled one-dimensional anisotropic oscillators and when the GWs pass through, they will weakly interact with GW. For example, each arm of LIGO apparatus can be thought as a point particle which is oscillating with time dependent frequency in two independent directions. When a GW passes by, due to the quadrupolar nature it creates oscillations in the plane perpendicular to its motion. Thus the effective dynamics of the interaction of GWs with the detector is a planar problem. The interaction of linearized GWs with our detector system is being then manifested through a particular quantal geometric phase-shift in the quantum states of the oscillators. Here we provide the estimation of this phase shift. Thus we hope that visualization of the effects of this BP on various physical phenomenon can be potential candidate to know about LFGWs.

In fact, the universal appeal of the quantal BP can be appreciated from the variety of contexts in which it has surfaced such as the Born -Oppenheimer approximation in molecular physics, fractional statistics, anomalies in gauge field theories, quantum Hall effect and several other situations \cite{forte,resta,dts,niu,natureqhe}. Moreover, BP comes in great accordance with the famous Unruh effect in the Unruh-DeWitt detectors, as presence of Berry phase in a version of Unruh-DeWitt detectors can serve as a direct consequence of the Unruh effect \cite{mann}. This phase, if detected, will lead to an indirect observation of the Unruh radiation. In this paper, we intend to show the footprint of the GWs on the quantum detectors. The gravitational counterpart of this geometric phase is indicative of the deflection of the detector's trajectory on account of the passing of GWs \cite{relicgravi}. Thus, a study of emergent BP has its interesting interpretations and consequences.

The organization of this article is as follows. In Section \ref{Sec2}, we provide the quantum vibrating detector model with anisotropic time-dependent frequencies interacting with low frequency mode of GWs. The computation of the BP has been presented in Section \ref{Sec3}. Section \ref{Sec4} concludes the paper.  We also provide six appendices to present the detailed steps of the calculations and supporting analysis. In the first Appendix A,  we provide a brief overview of linearized gravitational waves. We then show in Appendix \ref{AppB} how it is possible to construct a Hamiltonian which is equivalent to one we start with. This facilitates the subsequent computation. A brief derivation of BP in the Heisenberg picture is then presented in Appendix \ref{AppC}. 
In Appendix \ref{AppD}, we provide the BP's derivation based in the Schrodinger picture. Finally, we demonstrate an explicit computation of the BP and its variations with respect to the detector frequency amplitude in Appendix \ref{AppE}.

\section{Vibrating detector model}\label{Sec2}
For the linearized version of Einstein's theory of gravity, it is observed that the separation of geodesics, perpendicular to the direction of GW propagation, satisfies a very simple relation $(d^2\Delta x^{i})/dt^2 = - R^{i}_{~0j0}\Delta x^j$ \cite{MTW,maggiore} (see also a brief discussion in Appendix \ref{AppA}). This can be considered as the two dimensional motion of a particle (hence the spatial indices $i,j = 1,2$), influenced by GWs, relative to a fixed reference point under the forcing term given by $- R^{i}_{~0j0}\Delta x^j$. Now if we consider a detector, like LIGO, then the ends points of its each arm can be taken as a point particle. In this scenario each of the arms will follow two dimensional motion on a plane perpendicular to the direction of GW propagation which is driven by this equation of motion. For our case we keep this detector under an influence of another given force $F^i$ (non-geometric), the explicit form will be mentioned later.

Under this model end points of each arm will be driven by the following equation of motion $m\ddot{x}^l=-m R^l_{~0k0} x^k+F^l$,
where for brevity, $\Delta x$'s is being denoted by $x$'s by considering a fiducial fixed (reference) origin. Here, $m$ is the mass of the particle (e.g. end point of the arm). 
Now using $R^j_{~0k0}=\partial_t\Gamma^j_{k0}$, the Lagrangian corresponding to the above equation is $L'=\sum_j(\frac{1}{2}m\dot {x^{j}}^2+\frac{m}{2}\sum_kx^jx^k{\partial_t\Gamma^j_{0k}}-V_j)$ which upto a total derivative term can be taken as
\begin{equation}
L=\sum_j(\frac{1}{2}m\dot {x^{j}}^2-m\sum_k\Gamma^j_{0k}\dot{x^j}x^k-V_j)~,
\label{Lagrange2}
\end{equation}
where $V_j$ represents the external potential corresponding to the force $F^j$. 
The canonical Hamiltonian for (\ref{Lagrange2}) at the linearized order is then
 \begin{equation}
 H=\sum_j(\frac{p_j^2}{2m}+\sum_k\Gamma^j_{0k}x^kp_j + V_j (x^a))~.
 \label{H}
 \end{equation}
This Hamiltonian, written in slightly different manner, was recently considered in \cite{PWZ2} to probe the quantum nature of gravity in a two-particle detector model. In fact a similar model, introduced earlier in \cite{Ref1}, has also been employed in the context of noncommutative quantum mechanics (see e.g. \cite{Ref2}) for a different purpose.
 
The GW is expressed as $h_{jk}=2\chi(t)(\epsilon_{\times}\sigma_{1jk}+\epsilon_{+}\sigma_{3jk})$ \cite{MTW,maggiore}. Here $2\chi(t)$ is the amplitude of the GW and $\sigma_{1jk}$ is the $(jk)^{th}$ element of the Pauli matrix $\sigma_1$ and so on. Then the second term in (\ref{H}) will provide a term like $\sim (x_1p_2+x_2p_1)$ which corresponds to mutual interaction between the two directions of the single arm through GWs. In order to simplify the future calculation, it is customary to work on those phase-space variables in which such cross-terms can be eliminated. This can be done using unitary transformations $\tilde{x_i} = U_{ij} x_j$, $\tilde{p_i} = U_{ij}p_j$ with $U = e^{-\frac{i\theta\sigma_2}{2}}$. For our model, we take $V({\tilde{x}^a}) = \frac{1}{2}m\sum_j \Omega_j^2(t)\tilde{x}_j^2$ and then we have the total Hamiltonian in hermitian form as (see  Appendix \ref{AppB}),
\begin{eqnarray}
\label{workH}
H&=&\sum_{i=1,2} \Big(\alpha \tilde{p}_i^2+\beta_i\tilde{x}_i^2\Big)+\gamma(\tilde{x}_1\tilde{p}_1+\tilde{p}_1\tilde{x}_1)
\nonumber
\\
&-&\gamma(\tilde{x}_2\tilde{p}_2+\tilde{p}_2\tilde{x}_2)~,
\end{eqnarray}
where $\alpha=\frac{1}{2m}, \beta_j=\frac{1}{2}m\Omega_{j}^2$ and $\gamma=\dot{\chi}(t)\tilde{\epsilon}_{+}$. Here we have $\tilde{\epsilon}_+ = \epsilon_+ \cos\theta + \epsilon_{\times}\sin\theta$, with $\tan\theta = \frac{\epsilon_{\times}}{\epsilon_+}$.
Note that Eq. (\ref{workH}) represents the Hamiltonian for two anisotropic one-dimensional oscillators, each interacting independently with gravitational waves (GWs). Mutual interaction among the oscillators has been avoided by these choices to investigate the sole effect of GWs. This scenario can be understood as follows. Initially, the endpoints of one of the arms of LIGO are undergoing anisotropic oscillations in two perpendicular directions. When a GW passes through these arms, both $\Omega_{1,2}(t)$ in the potential, as described above, take the form of slowly varying periodic functions of time. Their time periods are finely adjusted to match the frequency of the incoming low-frequency gravitational wave (LFGW) mode. This adjustment results in the Hamiltonian (\ref{workH}), which exhibits periodicity with a period of $T = \frac{2\pi}{\omega_{g}}$. The choice of making $\Omega_{i}$ time-dependent and anisotropic for calculating Berry phases will be elaborated on in the next subsection.


 
Just for completeness, it may be mentioned that the form of $V$ in (\ref{H}) as a function of original coordinates can be found by applying the reverse unitary transformation. In this case this is given by  
\begin{eqnarray}
&&\sum_{j}V_{j}(x_{1},x_{2};t)=\frac{1}{4}m(\Omega^{2}_{1}+\Omega^{2}_{2})(x^{2}_{1}+x^{2}_{2})+ 
\nonumber 
\\
&+&\frac{1}{4}m\epsilon_{+}(\Omega^{2}_{1}-\Omega^{2}_{2})(x^{2}_{1}-x^{2}_{2})
\nonumber
\\
&+& \frac{1}{2}m\epsilon_{\times}(\Omega^{2}_{2}-\Omega^{2}_{1}) x_{1}x_{2}~.
\label{PRD1}
\end{eqnarray}
This structure of the potential indicates coupling between the harmonic oscillator modes, with the strength of coupling determined by $\epsilon_\times$. Furthermore, the choice of oscillation frequencies is contingent on the value of $\epsilon_{+}$. Notably, this type of potential has previously been employed in the study of gravity-induced entanglement, as discussed in \cite{Sup,kaku}. However we will work on tilde coordinates. This will not only simplify the analysis, but also such a choice incorporates only the interaction among the individual oscillators and GWs while the intra-interaction between them does not appear.

 It's worth highlighting that our choice to synchronize the time periods of the detector's frequency parameters with the low-frequency gravitational wave's frequency holds significant importance. This synchronization is critical as it requires a system Hamiltonian involving multiple time-dependent parameters with same time period to induce a nontrivial adiabatic Berry's phase shift in the quantum detector states \cite{MVB1}. Consequently, low-frequency gravitational waves naturally trigger an adiabatic evolution in the adjustable oscillator detector. This alignment is crucial for preserving the cyclicity condition of the Hamiltonian, which guarantees that a set of parameters, varied through a closed path (C) and subsequently returned to their original values, complies with the principles of the traditional adiabatic theorem. Ultimately, this alignment paves the way for the emergence of a nontrivial Berry's geometric phase.

Note that (\ref{workH}) can be rewritten in terms of the generators of $SU(1,1)$ group:
\begin{equation}
H=\alpha (T_1^{(1)}+T_1^{(2)})+\sum_{i=1,2}\beta_i T_2^{(i)}+\gamma (T_3^{(1)} -T_3^{(2)} )~,
\end{equation}
where $T^{(i)}_1=\tilde{p}_i^2, T^{(i)}_2=\tilde{x}_i^2, T^{(i)}_3=\tilde{x}_i\tilde{p}_i+\tilde{p}_i\tilde{x}_i$ are the three Lie algebra elements of $SU(1,1)$. It is a direct sum of two independent $SU(1,1)$ algebras corresponding to the two oscillator modes. The geometry associated with the parameter space of $SU(1,1)$ is traversed by the state vector cyclically, which shows up  a geometric phase shift after the complete cycle (see \cite{BG} for more details). Therefore the corresponding states must aquire BP. We will now calculate this.

\section{Computation of Berry phase}\label{Sec3}
In order to perform the quantum mechanics we define two ladder operators:
\begin{equation}
a_{1,2}=A_{1,2}(t)\Big[\tilde{p}_{1,2}+C_{1,2}(t)\tilde{x}_{1,2}\Big]~;
\end{equation}
such that only non-vanishing one is $[a_i,a_i^\dagger] = 1$ with, $A_i=\sqrt{\frac{1}{2m\hbar\omega_i}}, C_{1,2}=\frac{1}{\alpha}(\pm \gamma-\frac{i\omega_{1,2}}{2})$ and $\omega_i=\sqrt{\Omega_i^2-4\gamma^2}~>0$. Positive sign is for $C_1$ and other one is for $C_2$.
These two, along with their adjoints, readily diagonalize (\ref{workH}) as,
 \begin{equation}
	H=\hbar\sum_j(\omega_j{a_j}^\dagger a_j)+\frac{\hbar}{2}(\omega_1+\omega_2)~.
 \label{hm}
\end{equation}

 The time evolution of these operators are determined from the Heisenberg equation of motion.
This yields
\begin{equation}\label{ladderevo1}
\dot {a_1}=[M_1-\eta_1] a_1+\eta_1 a_1^\dagger~,
\end{equation}
where,  $M_1=-i\omega_1+\frac{\dot{A}_1}{A_1}$ and $\eta_1=\frac{\dot{{ C_1}}}{2im\omega_1}$.
The same for $\dot a_1^{\dagger}$ is obtained by taking the hermitian conjugate of (\ref{ladderevo1}).
Note that $\gamma$ is related to the GWs and so can be regarded as a time-dependent parameter, which is taken to be varying adiabatically. To quantify the adiabaticity, let us define a dimensionless parameter $\epsilon$ as
\begin{equation}
\epsilon=\frac{T_{i}}{T_{e}}\sim\frac{\omega_{g}}{\omega_{n_1,n_2}}<<1~.
\label{epsilon}
\end{equation}
Here, $T_{i}=\frac{\hbar}{E_{n_1,n_2}}\sim \omega^{-1}_{n_1,n_2}$ represents the internal time scale, where $\omega_{n_1,n_2}=(n_{1}+\frac{1}{2})\omega_{1}+(n_{2}+\frac{1}{2})\omega_{2}$ corresponds to the instantaneous frequency associated with the non-degenerate energy level $E_{n_1,n_2}=\hbar\omega_{n_{1},n_{2}}$, characterized by the quantum numbers $n_1$ and $n_2$, of the system Hamiltonian (\ref{hm}). On the other hand, the term $T_{e}=|\frac{\bra{n_1,n_2}\frac{\partial {H}(t)}{\partial t}\ket{f}}{E_{n_1,n_2}-E_f}|^{-1}\sim \omega^{-1}_{g}$ characterizes the external time scale. This is because the parameter space of the system Hamiltonian depends on a periodic function of time with a periodicity that depends on $\omega^{-1}_{g}$, as mentioned earlier. In this context, $\omega_{g}$ represents the frequency of external gravitational wave perturbations.
The parameter $\epsilon$ quantifies how slowly the external perturbation changes the system Hamiltonian compared to the energy gap between the initial quantum states, defined by the quantum numbers $n_1$ and $n_2$, and other final states represented by $\ket{f}$.

Under the adiabatic approximation, we consider $\gamma(t)$ and $\Omega_{i}(t)$ as slowly varying periodic parameters. As a result, we retain their first-order derivatives (representing the first order in adiabaticity) while neglecting higher-order derivatives (higher adiabaticity). Furthermore, we do not take into account terms that involve the square of their first derivatives. In this situation combination of (\ref{ladderevo1}) and that for $a^{\dagger}_{1}$, under adiabatic approximation yields
\begin{equation}
\ddot{a}_1=(M_1-\eta_1)\dot a_1-i\dot\omega_1a_1+\eta_1\tilde M_1a_1^\dagger~.
\end{equation}
Finally, eliminating ${a}_1^\dagger$ by using (\ref{ladderevo1}) one obtains a linear second-order differential equation for $a_1$ as,
\begin{equation}
\ddot a_1=\bigg(2\frac{\dot A_1}{A_1}+i\frac{\dot{C_1}}{{2m\omega_1}}\bigg)\dot a_1-\bigg(\omega_1^2+i(\dot\omega_1-\eta_1\omega_1)\bigg)a_1~.
\label{BRM1}
\end{equation}

The solution of the above one can be obtained using WKB like trick. Consider the following ansatz:
\begin{equation}  a_1(t)=\rho(t) e^{\frac{1}{2}\int d\tau [i\frac{\dot{C_1}}{{2m\omega_1}}+\frac{2\dot A_1}{A_1}]}~,
\label{WKB}
\end{equation}
where the time-dependent function $\rho(t)$ has to be determined. Then a detailed calculation yields the solution as  (see Appendix \ref{AppC} for details)
\begin{equation} 
a_1(T)=a_1(0)e^{-i\int_0^T (\omega_1-\frac{\dot{\gamma}(\tau) }{\omega_{1}(\tau)})d\tau}~.
\end{equation}
This suggests that apart from the usual dynamical phase factor of $e^{-i\int _0^T \omega_1 d\tau}$, the system develops an additional geometric phase given by,
\begin{equation}\label{finalres}
\phi_{g}^{(1)}=\int_0^T \frac{\dot\gamma}{\sqrt{(\Omega_1^2-4\gamma^2)}}d\tau~.
\end{equation}
Similarly, on studying the evolution of the second mode $a_2$, the BP obtained is given by:
\begin{equation}\label{finalres2}
\phi_{g}^{(2)}=-\int_0^T \frac{\dot\gamma}{\sqrt{(\Omega_2^2-4\gamma^2)}}d\tau~.
\end{equation}
Note the appearance of the overall negative sign here in contrast to (\ref{finalres}) as can be anticipated from the structure of the Hamiltonian (\ref{workH}).

Before we proceed further, let us pause for a while and
make some pertinent comments.

(i) It's important to recognize that our internal time scale ($T_{i}$) is intimately related to the instantaneous normal mode frequencies $\omega_{1}(t)$ and $\omega_{2}(t)$ of the system Hamiltonian (\ref{hm}). Specifically, we have $\omega_{i}(t)=\sqrt{\Omega^{2}_{i}(t)-\gamma^{2}(t)}$. So, it becomes clear from the expression of $\omega_{i}$ that $\Omega_{i}$ indeed contributes to the determination of the system's internal time scale, $T_{i}$. Notably, in the absence of any gravitational wave perturbations, the primary responsibility for defining this time scale falls upon $\Omega_{i}$.
Furthermore, the time-dependent behavior of the frequencies $\Omega_{i}(t)$ and the parameter $\gamma(t)$ in our mechanical oscillators is of significant importance as it introduces an additional time scale to the dynamical system. This additional time scale is referred to as the external time $T_{e}$, as defined previously. $T_{e}$ determines the rate at which the system's parameters change . When we mention adiabaticity, we are essentially emphasizing that $T_{e}$ is significantly greater than the internal time scale $T_{i}$ (or that $\epsilon<< 1$ , as mentioned earlier). This condition implies that the system's parameters change slowly compared to the internal dynamics of the system. Consequently, it prevents the system from making abrupt transitions to different, non-degenerate states.

(ii) From the expression of the BP that emerges in Eq. (\ref{finalres}) and Eq. (\ref{finalres2}), it can be noted that the extra phase will be an integral of exact differential thus becoming zero over a complete cycle if the oscillator frequencies are taken to be just constants, not time-dependent ($\phi^{(i)}=\frac{1}{2}\oint d(\sin^{-1}\frac{2\gamma}{\Omega_i})$). Therefore, it is crucial to consider these frequencies as time-dependent ones. From the standpoint of differential geometry \cite{sp,Ref4}, the geometric significance of the Berry phase becomes nontrivial when the integral of the one-form (the phase integral) is a closed but not exact form. This condition highlights the importance of time-dependent oscillator frequencies in capturing nontrivial geometric effects associated with the Berry phase.
On the other hand, since we have previously observed that our time-dependent system Hamiltonian is an algebraic element of the $SU(1,1)$ Lie group (expressible as a linear combination of $SU(1,1)$ group generators \cite{jp}), the emergence of the Berry phase can be attributed in our case to the breaking of time-reversal symmetry in the Hamiltonian due to the presence of a generator explicitly breaking this symmetry at the instantaneous level \cite{jm}. Furthermore, the parameter space of the system Hamiltonian can be identified with the parameter space of the $SU(1,1)$ group manifold. To obtain a nontrivial geometric phase shift, a set (of at least two) parameters, including the time-dependent coefficient of the time-reversal symmetry-breaking term, must be varied adiabatically to form a closed loop ``$C$" in the parameter space (see details in the reference \cite{parthab,parthac}).
Therefore, it is a common
wisdom that for a non-trivial Berry's geometrical phase to exist, the Hamiltonian must possess more than one time-dependent parameter, allowing the state vector to exhibit anholonomy when transported around a closed loop $C$ adiabatically in the corresponding parameter space. In contrast, the presence of only one time-dependent parameter causes the closed loop to be trivial (effectively collapsing to a one-dimensional line), making it contract to a point in the parameter space, resulting in a vanishing geometric phase. This rationale justifies our consideration of the $\Omega_i$'s as time-dependent.

Furthermore, if the frequencies of the oscillators are assumed to be zero, it would render the resulting system non-oscillatory and purely damped, hence unstable, with no lower bound for the energy. More importantly, from a practical standpoint, our model detector closely adheres to Weber's initial concept of mechanical resonant bar detectors for gravitational wave detection \cite{weber,weber2}.


(iii) In the context of time-dependent systems, the occurrence of level crossings is significant. Level crossing happens at a specific moment when the time-dependent parameters of the Hamiltonian reach values such that, for a particular pair of non-degenerate states, $E_{n_{1},n_{2}}(t)=E_{m_{1},m_{2}}(t)$ with $(n_{1},n_{2})\neq (m_{1},m_{2})$. In standard quantum mechanics, the proof of the adiabatic theorem asserts that during the evolution of a system in parameter space, there should be no level crossings. This theorem ensures that as one traces the curve in the parameter space defining the Hamiltonian from $H_i$ to $H_f$, an $n$-th eigenstate under the initial Hamiltonian $H{(t_{i}=0)}$ is adiabatically transported to the $n$-th eigenstate under the final Hamiltonian $H(t_{f}=T)$, provided the system changes gradually. Indeed, this theorem is based on the assumptions of a discrete and non-degenerate spectrum, as long as it is ensured that the trajectories of two eigenvalues do not intersect \cite{ze,rb,ke}.
Additionally, the occurrence of level crossings can lead to non-adiabatic transitions, which in turn introduce complexity into the system's behavior, as discussed in \cite{e,ek}. In our specific problem, it's worth noting that we do not encounter level crossing even when the oscillator frequencies $\Omega_i(t)$ are equal. Nonetheless, when we introduce anisotropy to intentionally break the rotational symmetry within these oscillator frequencies, we can avoid the non-abelian characteristics of the BP, as discussed in \cite{zee,gb,Ref3}. Then the system continues to support a discrete, non-degenerate spectrum throughout its time evolution.

(iv) Furthermore, it is also crucial to emphasize the absolute necessity of non-vanishing denominators ($\omega_{i}(t)>0$) in the integrands during the adiabatic variation over the period $T$. This condition is absolutely essential, as our entire phase derivation relies on the adiabatic approximation, which must hold throughout the system's evolution. Should the denominator reach zero (i.e. $\omega_i=0$) at any point during this evolution, it would result in a catastrophic breakdown of the adiabatic theorem at that particular point (see Eq. (\ref{epsilon})).  
This underscores the need for continuous non-zero denominators to maintain the adiabatic theorem's integrity (see detailed analysis in chapter $XVII$ of \cite{ze} as well as \cite{rb}). Therefore, physically, it is natural to assume that $\omega_i$ is always positive at all times. This condition ensures that the integral form of the Berry phase is always well-defined.

Now returning back to our main objective. An important aspect of BP is its geometrical nature. This will be more transparent when expressed as the integral of a 1-form along a closed circuit within the parameter space:
\begin{equation}
	\phi^{(i)}_{g}[C]= (-1)^{i+1} \oint_C\frac{1}{\omega_i}\nabla_R \gamma\cdot dR; ~~i=1,2
\end{equation}
where $R$ is a vector in the space of parameters and the Hamiltonian changes via the parameters in such a manner that it makes a closed circuit in the space of parameters where it returns to its initial value after a cycle. Thus this additional phase is a functional of the circuit traversed in the parameter space and is manifestly independent of how the path has been traversed. 

We obtained this geometric phase shift in Heisenberg picture; but it can be readily obtained in the more familiar form of BP acquired by state vectors. For that we revert back to the Schrodinger picture and after a straightforward calculation (see Appendix \ref{AppD}), we have the geometric phase acquired by an arbitrary Fock state $\left|n_{1},n_{2};R(t=0)\right\rangle _{H.O.+GWs}$ to be given by (also see \cite{parthab} for details)
\begin{equation}
	\phi_{B}^{(n_{1},n_{2})}=\phi_{B}^{(0,0)}+n_{1}\phi^{(1)}_{g}+n_{2}\phi^{(2)}_{g}
\end{equation}
and the total phase as,
\begin{eqnarray}
&&\Phi^{(n_1,n_2)}=\Phi^{(0,0)}+\bigg[n_{1}(\theta^{(1)}_{d}+\phi^{(1)}_{g}) 
\nonumber 
\\
+&&n_{2}(\theta^{(2)}_{d}+\phi^{(2)}_{g})\bigg]~,
\end{eqnarray}
where $n_1,n_2$ are semi-positive definite integers representing the eigenvalues of the number operators $a_1^{\dagger}a_1$ and $a_2^{\dagger}a_2$ respectively. In the above the dynamical part of the phase is $\theta^{(i)}_{d}=\int_{0}^{T}\omega_i(\tau)d\tau$.
Note that it is the difference of the BPs of different eigenstates, which contributes in the expectation value of any operator at time $t$ in a state obtained from any
initial state and evolving under an adiabatic Hamiltonian,
where the ground state contribution $\phi_B^{(0,0)}$ cancels out. This idea also resonates while carrying out experiments concerning measurement of BP. 

The obtained phase (\ref{finalres}) has some interesting characteristics. Firstly, both kinds of polarizations $\epsilon_{+},\epsilon_{\times}$ contribute to BP. Secondly, it might seem that the phase, containing second-order time derivative of the time-dependent GW amplitude (as $\gamma\sim \dot{\chi}$), is negligible. However, in such a consideration, the interaction part of the Lagrangian $L'$ would have vanished, which cannot be true. Therefore, this geometric phase would be consistently observed by tuning the external frequency $\Omega_1$ or $\Omega_2$. A characteristic analysis of estimated BP as a function of oscillator frequency amplitude shows that the  
BP monotonically decreases with the increase of amplitude (see Appendix \ref{AppE}).

\section{Discussion}\label{Sec4}
Firstly, we summarize our findings. We have considered a two-dimensional time-dependent anisotropic harmonic oscillator detector to probe the passing of GWs. With a suitable rearrangement of the terms, we can show that such a system is reminiscent of  a generalised harmonic oscillator along with a boost term in phase space. Thereafter, we have performed a proper redefinition of the phase-space variables to eliminate the boost term which facilitates our subsequent analysis smoothly. At the end, we computed the BP in the Heisenberg picture and found that both plus and cross polarization modes are responsible for the existence of the phase. In other words, this additional phase in the detector's wave function is due to the coupling of the detector with the GWs. Whereas in absence of GWs, there is only the dynamical phase. It will be worth-mentioning here that there exists BP exhibiting Hamiltonians whose BP may be removed by a suitable time-dependent canonical transformation \cite{liang}. However, in such a case, the BP reappears in the dynamical part retaining its geometric nature. In our case too, the Hamiltonians corresponding to the Lagrangians $L'_j$ and (\ref{Lagrange2}), being connected by a time-dependent canonical transformation, lead to the same expression for this additional geometric phase over and above the trivial dynamical phase. In-fact, our approach does not follow the one  used in modelling the Weber detectors \cite{weber,weber2} but instead we consider an equivalent and perhaps more illuminating form of the interaction in order to compute the BP as has been also recently considered in \cite{PWZ2}, and this choice of the system  Hamiltonian has been further motivated by its somewhat resemblance with that of the problem of a charged particle moving in two dimensions in an applied magnetic field acting perpendicular to the plane of motion. But, as we just stated, the choice of the Hamiltonians, which are related by time-dependent canonical transformations, has no effect on how BP is expressed because this additional phase is invariant under both unitary and gauge transformations \cite{liang}.
Furthermore, the introduction of an explicitly broken time reversal symmetry, achieved through the inclusion of a dilatation term, plays a pivotal role in the generation of non-vanishing BPs within the oscillator detector. The passing gravitational wave possesses a quadrupole nature, leading to the induction of two-mode squeezing in the oscillator detector.

In our methodology, BPs are determined by solving the evolution equations for $a_{i}(t)$ in the Heisenberg picture. Our motivation for this choice primarily stems from the inherent relationship between the ladder operators of the quantum system and analogous operators resembling number ($\hat{N}$) and phase ($\hat{\theta}$): $a_{i}=\sqrt{\hat{N}}e^{\hat{i\theta}}$, as elaborated in reference \cite{D}. Consequently, this approach serves as a natural framework for exploring additional phase factors beyond the dynamical phase throughout the adiabatic evolution of the system's Hamiltonian. In addition, our method simplifies the systematic identification of the associated classical Hannay angle \cite{H} (see Appendix E).

Furthermore, it may be noted that as the frequency of the oscillator detector sets a scale in the system, tuning it to a range of a few Hertz will enable to detect GWs of considerably lesser frequencies as the adiabatic condition implies the slower variation of the perturbing gravitational influence for the existence of the geometric phase. This suggests that one would be, at least in ideal situations, able to detect GWs of frequencies less than a few Hertz from this geometric phase shift in the detector's states.
On the other hand, this demonstrated BP whether leads to an entanglement in the quantum detector's degrees of freedom is an important and intriguing question that we want to address in the near future \cite{rov1,rov2}, which will be a step towards probing the quantum nature of gravitational waves through quantum-mechanical detectors. 

As a final remark, the emergent nature of GW-induced BP may be detectable in principle, but we
are still far from providing a quantitative measurement of this phase.
The detectability of this phase may therefore serve as a new probe of very weak gravitational waves. A theoretical aspect of detecting the weak GW-induced
BP may be explored in a squeezed state formalism \cite{albrecht}, and the geometric phase may be detectable
from the phase difference in a suitably designed interference experiment. In fact, a scheme for detecting harmonic oscillator's BP through the vibrational degree of freedom of trapped ions has been laid out in \cite{bose2000}, and it may be extended for the
generalized harmonic oscillator model. We are working on it and hope to return to some of these issues in a future work soon. 

\vskip 3mm
\noindent
\underline{Acknowledgements}: Two of us, P.N. and S.K.P., express their gratitude towards Biswajit Chakraborty for a critical reading of the manuscript and for his constant encouragements. P.N. thanks A.P. Balachandran, V.P. Nair , A. P. Polychronakos and  Frederik G. Scholtz for their insightful and critical comments on the work. Also, P.N. thanks M. Berry for a useful email correspondence. Finally, P.N. and S.K.P. thankfully acknowledge the visiting research fellowships provided by the S.N. Bose National Centre during the period when this work was initiated. PN also would like to acknowledge the generous support from the Institute of Theoretical Physics, Stellenbosch University, in the form of postdoctoral fellowship. Finally, we would like to thank the referee for providing valuable and constructive comments on the earlier version of the draft. 





\begin{widetext}

\begin{appendix}
\section*{Appendices}


\section{Basic Review of Linearized Gravity}\label{AppA}
\renewcommand{\theequation}{A.\arabic{equation}}

Einstein's theory of general relativity is of great success in classical GR. It can almost accurately describe all the phenomena at larger mass scales. Now, the phenomenon of gravitational waves emerges from a linearized approximation of Einstein's GR where small perturbations are considered over the usual Minkowski flat space-time:
\begin{equation}
g_{\mu\nu}=\eta_{\mu\nu}+h_{\mu\nu} 
\label{g}
\end{equation}
with, $|h_{\mu\nu}|\ll 1$. The Christoffel connection coefficients and the Riemann curvature tensor in this case then take the following forms:
\begin{eqnarray}
	&&\Gamma^{\mu}_{~\nu\sigma}=\frac{1}{2}\eta^{\mu\rho}(\partial_{\sigma} h_{\nu\rho}+\partial_{\nu} h_{\sigma\rho}-\partial_{\rho}h_{\nu\sigma})~;
	\\
&&R^{\mu}_{~\rho\sigma\nu}=\frac{1}{2}{\eta^{\mu\lambda}}(\partial_{\sigma}\partial{\rho}h_{\nu\lambda}-\partial_{\sigma}\partial_{\lambda}h_{\nu\rho}-\partial_{\nu}\partial_{\rho}h_{\sigma\lambda}+\partial_{\nu}\partial_{\lambda}h_{\sigma\rho})~.
\label{riemann} 
 \end{eqnarray}
On extremizing the Einstein-Hilbert action for this case
\begin{equation} S_{E-H}=\frac{1}{16\pi G}\int d^4x \bigg( \sqrt{-g}R+\mathcal{L}_{matter}\bigg)~,
 \end{equation}
yields the linearized version of Einstein's equation
\begin{equation} 
	\Box \bar{h}_{\mu\nu}=-16\pi GT_{\mu\nu}~~;~\bar h_{\mu\nu}=h_{\mu\nu} -\frac{1}{2}\eta_{\mu\nu}h~,
\end{equation}
in terms of the trace-reversed perturbation $\bar{h}_{\mu\nu}$.
Thus in regions outside of the sources, one has 
\begin{equation}
\Box \bar h_{\mu\nu}=0~;
\end{equation}
whose solutions are basically the gravitational waves
\begin{equation} 
\bar h_{\mu\nu}=Re(\epsilon_{\mu\nu}e^{ik_{\rho}x^{\rho}})~.
\end{equation} 
Here $\epsilon_{\mu\nu}$ is some complex, symmetric polarization matrix and $k^{\mu}$ is a real wave-vector.
Via the transverse-traceless (TT) gauge condition $h_{0\mu}=0$, $h^{\mu}_{\mu}=0$ and $\partial^ih_{ij}=0$, we can completely fix the polarisation matrix as \cite{MTW,maggiore},
\begin{equation}\epsilon_{\mu\nu}=
	\begin{pmatrix}
		0 & 0 & 0 & 0\\
		0 &  \epsilon_+ &  \epsilon_{\times} & 0\\
		0 &  \epsilon_{\times} & - \epsilon_+ & 0\\
		0 & 0 & 0 & 0
	\end{pmatrix} \label{polari}
\end{equation}
where $\epsilon_+$ and $\epsilon_{\times}$ correspond to plus and cross polarizations of the gravitational waves respectively.
Let us consider two nearby geodesics $x^{\mu}(\tau)$ and $x^{\mu}(\tau)+\Delta x^{\mu}(\tau)$ in the background (\ref{g}). Then the equation of motion of the separation vector $\Delta x^{\mu}(\tau) $ is given by \cite{MTW,maggiore}, 
\begin{equation}
	\frac{D^2\Delta x^{\mu}}{D\tau^2}=-R^{\mu}_{\nu\rho\sigma}\Delta x^{\rho}\frac{dx^{\nu}}{d\tau}\frac{dx^{\sigma}}{d\tau}
\end{equation}
where, $\frac{D}{D\tau}$ is the covariant derivative with respect to the proper time $\tau$. We now choose the proper detector frame to set up and study our laboratory detector physics. Using the definition of Riemann tensor for linearised theory (\ref{riemann}), the above equation simplifies to \cite{MTW,maggiore},
\begin{equation}\label{properdetector1}
	\frac{d^2\Delta x^{i}}{dt^2}=-R^{i}_{0j0}\Delta x^{j}
\end{equation}
where the coordinate time $t$ can be approximated to be the proper time $\tau$ upto the first order in perturbations. Also note that here, the indices $i,j$ take values only 1 and 2, owing to (\ref{polari}).
Further implementing the TT gauge conditions, we can rewrite the above equation as,
 \begin{equation}\label{properdetector2}
 	\frac{d^2\Delta x^{i}}{dt^2}=\frac{1}{2}\frac{d^2h^{i}_{j}}{dt^2}\Delta x^{j}~.
 \end{equation}
Clearly, the above is a Newtonian description i.e. non-relativistic equation of motion. Typically in the present paper, this is the physical situation of most interest to us in order to model the dynamics of a non-relativistic detector and to study the consequences of the passing of GWs in the ambient space-time.

\section{Unitary equivalent Hamiltonian (\ref{workH})}\label{AppB}
\renewcommand{\theequation}{B.\arabic{equation}}

In TT gauge, the relative motion of two free falling particles on the gravitational wavefront (propagating along $z$ direction) can be described in terms of the following system Hamiltonian as
\begin{equation}
    \hat{H}_0=\sum_i\frac{\hat{p}^{2}_{i}}{2m}+ \sum_{j,k}(\frac{\Gamma^j_{0k}}{2}(\hat{x}^k\hat{p}_j+\hat{p}_j\hat{x}^k))~; 
    \label{h}
\end{equation}
where, $i,j=1,2$ and the gravitational waves interaction coupling term $ \Gamma^j_{0k}$ is a $su(2)$ Lie algebra valued field:
\begin{equation}
\Gamma^j_{0k} =2\dot{\chi}(t)(\epsilon_{\times}\sigma_{1jk}+\epsilon_{+}\sigma_{3jk})~.
\end{equation}
Accordingly the system Hamiltonian (\ref{h}) may be rewritten as
\begin{equation}
    \hat{H}_0=\sum_j\frac{\hat{\tilde{p_j}}^{2}}{2m}+ \frac{\tilde{\Gamma}^1_{01}}{2} (\hat{\tilde{x_1}}\hat{\tilde{p_1}}+\hat{\tilde{p_1}}\hat{\tilde{x_1}}) +\frac{\tilde{\Gamma}^2_{02}}{2} (\hat{\tilde{x_2}}\hat{\tilde{p_2}}+\hat{\tilde{p_2}}\hat{\tilde{x_2}}),
    \label{k}
\end{equation}
after by applying a unitary ($su(2)$), albeit time independent, transformations:
\begin{equation}
\tilde{x_i} = U_{ij} x_j;~~~~~ \tilde{p_i} = U_{ij}p_j;~~~~~~\Gamma^j_{0k} \rightarrow \tilde{\Gamma}^j_{0k}= (U\Gamma U^{\dagger})^j_{0k}= 2\dot{\chi}(t) \tilde{\epsilon}_+ \sigma_{3jk}~.
\end{equation}
In the above $U=e^{-i\frac{\theta\sigma_{2}}{2}} $, and  $ \tilde{\epsilon}_+ = \epsilon_+ \cos\theta + \epsilon_{\times}\sin\theta$, with $\theta = \tan^{-1}(\frac{\epsilon_{\times}}{\epsilon_+})$.

\section{Derivation of the expression (\ref{finalres})}\label{AppC}
\renewcommand{\theequation}{C.\arabic{equation}}

Now using the WKB-like ansatz (\ref{WKB}) in (\ref{BRM1}) one finds that the time-dependent function $\rho(t)$ satisfies,
\begin{equation} 
\ddot\rho+(u+iv)\rho=0~,
\end{equation}
where $u=\omega_1^2$ and $v=(\dot \omega_1-\eta_1\omega_1)$.
In the WKB method, we write the solution as, 
\begin{equation}
\rho(t)=\frac{c_1}{\sqrt{\Xi}(t)}e^{\int_0^t(i\Xi(\tau)-\zeta({\tau}))d\tau}+\frac{c_2}{\sqrt\Xi(t)}e^{\int_0^t (-i\Xi(\tau)+\zeta({\tau}))d\tau}~,
\label{ed}
\end{equation}
where we essentially have, 
\begin{equation} 
\Xi(t)+i\zeta({t})=\sqrt{(u+iv)}~,
\end{equation}
 and $c_{1}$, $c_{2}$ are arbitrary coefficients that can be used to find the general solution of the above differential equation.
In our case, it's important to highlight that when we take into account the adiabatic changes in both $u$ and $v$, we reach the following result:
 \begin{equation}
     \Xi(t)\approx \sqrt{u}=\omega_{1};~~~\zeta(t)\approx\sqrt{\frac{v^{2}}{4u}}=\frac{\dot{\omega}_{1}}{2\omega_{1}}-\frac{\eta_{1}}{2}.
 \end{equation}

 We now consider the initial condition that the solution must satisfy: $\rho_{1}(t=0)=a_{1}(t=0)$. Notably, only the phase factor of the second term with the coefficient $c_{2}$ in the solution (\ref{ed}) contributes to the dynamical phase of $a_{1}$ with the correct sign. This will become evident as we calculate $a_{1}(T)$. Consequently, we set $c_{1}=0$ in (\ref{ed}), this boils down to
\begin{equation}
\rho(t)=\frac{c_2}{\sqrt\Xi(t)}e^{\int_0^t (-i\Xi(\tau)+\zeta({\tau}))d\tau}~.
\label{ed2}
\end{equation}
At this stage, by using the initial condition, we can express the arbitrary coefficient $c_{2}$ as:
 \begin{equation}
    c_{2}= \sqrt{\Xi}(t=0)~a_{1}(t=0)~. 
\end{equation}
Then we arrive at the following solution for $\rho(t)$:
\begin{equation} 
	 \rho(t)=\sqrt{\frac{\Xi(0)}{\Xi(t)}} a_{1}(t=0) e^{\int_0^t(-i\omega_1+\frac{\dot\omega_1}{2\omega_1}-\frac{\eta_1}{2})d\tau}~.
 \end{equation}
 Now, the periodicity of the parameters implies that $\sqrt{\xi(0)}=\sqrt{\xi(T)}$. Therefore, by applying (\ref{WKB}) and considering the cyclic evolution of the system in the parameter space, we can effectively neglect the term involving only the exact derivatives. This allows us to separate the corresponding dynamical and geometric phase shifts as
 \begin{equation}
    a_{1}(T)= a_{1}(0) e^{-i\int^{T}_{0} d\tau (\omega_{1}(\tau)-\frac{\dot{\gamma}(\tau)}{\omega_{1}(\tau)})}~.
    \label{lad}
\end{equation}  
Similarly, it can be shown that the time evolution equation of  $a_{2}$  is identical to the one for $a_{1}$, except that  $\gamma$ is replaced by $-\gamma$.  So, we get
\begin{equation}
    a_{2}(T)= a_{2}(0) e^{-i\int^{T}_{0} d\tau (\omega_{2}(\tau)+\frac{\dot{\gamma}(\tau)}{\omega_{1}(\tau)})}~. 
\end{equation}

Looking at the second phase factor in the expressions of both the annihilation (corresponding creation) operators $a_1$ in ($\ref{lad}$), the additional phase factor obtained by leading behavior for adiabatic transport around a closed loop $C$ in time $T$ can be identified with the Berry phase or geometric phase (more precisely geometric phase shift) in the Heisenberg picture.

\section{Schroedinger Picture for Berry phase}\label{AppD}
\renewcommand{\theequation}{D.\arabic{equation}}

As previously stated, the transition from the Heisenberg picture to the Schrödinger picture allows us to express our findings in a more conventional manner in terms of the phase acquired by the state vector (as shown in, for instance, \cite{BDR}). In this section, we illustrate how our approach, relying on ladder operators, aids in the computation of the Berry phase within the framework of the Schrödinger picture.

Let us start by considering the instantaneous eigenstates of our system Hamiltonian as
\begin{equation}\label{5.4}
\ket{n_1,n_2;t}=\frac{1}{\sqrt{n_1!n_2!}}\left((a_1^\dagger(t))^{n_1}|0;t\rangle_{1}\right)\otimes\left((a_2^\dagger(t))^{n_2}|0;t\rangle_{2}\right)~.
\end{equation}
Following Berry's original work \cite{berry1}, during the adiabatic evolution of the system's Hamiltonian in the Schroedinger picture, the time evolution of the instantaneous eigenvectors is described by
\begin{equation}
  \ket{n_{1},n_{2};t=0} \longrightarrow \ket{\Psi_{n_1,n_2}(T)}=e^{-\frac{i}{\hbar}\int^{T}_{0} dt ~E_{n_{1},n_{2}}(t) } e^{i\phi^{(n_{1},n_{2})}_{B}}  \ket{n_{1},n_{2};T}~,
  \label{1}
\end{equation}
with $E_{n_{1},n_{2}}=(n_{1}+\frac{1}{2})\hbar\omega_{1}+(n_{2}+\frac{1}{2})\hbar\omega_{2}$, and
\begin{equation}
    \phi^{(n_{1},n_{2})}_{B}=\int^{T}_{0} dt \bra{n_{1},n_{2},t} \Big[i(\partial_t)_1\otimes I_{2}+I_1\otimes i(\partial_t)_2\Big] \ket{n_{1},n_{2},t}=\int^{T}_{0} dt \Big[\bra{n_{1},t} i\frac{\partial}{\partial t} \ket{n_{1},t}_{1} + \bra{n_{2},t} i\frac{\partial}{\partial t} \ket{n_{2},t}_{2}\Big]~.
    \label{2}
\end{equation}
Consequently, the integrand's first term can be rewritten as
\begin{equation}
\bra{n_{1},t} i\frac{\partial}{\partial t} \ket{n_{1},t}= \bra{n_{1}-1,t} i\frac{\partial}{\partial t} \ket{n_{1}-1;t}+\frac{i}{\sqrt{n}} \bra{n_{1};t} \frac{\partial a^{\dagger}_{1}}{\partial t} \ket{n_{1}-1;t}~,
  \label{3}
\end{equation}
where we have used the following facts $a_{i}(t)\ket{n_{i}}_{i}=\sqrt{n_{i}} \ket{n_{i}-1}$, and $a^{\dagger}_{i}(t) \ket{n_{i}}_{i}=\sqrt{n_{i}+1} \ket{n_{i}+1}$.

Now on using Eq(\ref{ladderevo1}) we easily arrive at
\begin{equation}
    \frac{\partial a^{\dagger}_{1}}{\partial t}=\frac{\dot{A}_{1}}{A_{1}}a^{\dagger}_{1}-\bar{\eta}_{1}(a^{\dagger}_{1}-a_{1})~,
    \label{4}
\end{equation}
where $\eta_{1}=-i\frac{\dot{C}_{1}}{2m\omega_{1}}$ with $C_{1}=2m(\gamma-i\frac{\omega_{1}}{2})$. Here the bar quantities signify the complex conjugate of the respective terms.
Then (\ref{3}) becomes
\begin{multline}
\bra{n_{1},t} i\frac{\partial}{\partial t} \ket{n_{1},t}_{1}= \bra{n_{1}-1,t} i\frac{\partial}{\partial t} \ket{n_{1}-1;t}_{1}-i \bar{\eta}_{1}+i\frac{\dot{A}_{1}}{A_{1}}=
 \bra{0,t} i\frac{\partial}{\partial t} \ket{0;t}_{1}+ n_{1}\frac{\dot{\bar{C}}_{1}}{2m\omega_{1}}+i\frac{\dot{A}_{1}}{A_{1}}~.
     \label{5}
\end{multline}
Similarly it can be shown that
\begin{equation}
\bra{n_{2},t} i\frac{\partial}{\partial t} \ket{n_{2},t}_{2}=
     \bra{0,t} i\frac{\partial}{\partial t} \ket{0;t}_{2}+ n_{2}\frac{\dot{\bar{C}}_{2}}{2m\omega_{2}}+i\frac{\dot{A}_{2}}{A_{2}}~.
     \label{6}
\end{equation}
Substituting Eqs (\ref{5}) and (\ref{6}) in Eq(\ref{2}) we arrive at
\begin{equation}
 \phi^{n_{1},n_{2}}_{B}=\phi^{(0,0)}_{B}+n_{1}\phi^{(1)}_{g}+n_{2}\phi^{(2)}_{g}~,
 \label{l}
\end{equation}
with $\phi^{(0,0)}_{B}=\int^{T}_{0}dt~ \bra{0,0;t} i\frac{\partial}{\partial t} \ket{0,0;t}$ and 
\begin{equation}
    \phi^{(1)}_{g}= \int^{T}_{0} dt ~ \frac{\dot{\gamma}(t)}{\omega_{1}(t)};~~~\phi^{(2)}_{g}=- \int^{T}_{0} dt ~\frac{\dot{\gamma}(t)}{\omega_{2}(t)}~.
    \label{bg}
\end{equation}
It's important to note that Eq. (\ref{l}) doesn't provide any additional information about $\phi^{(0,0)}_{B}$, as the choice of relative phases remains arbitrary. To simplify our analysis, we adopt a convenient phase choice that ensures the vanishing of $\phi^{(0,0)}_{B}$ as $\gamma$ becomes a constant and introduces a `zero-point' contribution \cite{DR} to Eq. (\ref{l}) as
\begin{equation}
\bra{0,0;t} i\frac{\partial}{\partial t} \ket{0,0;t}=\frac{\dot{\gamma}(t)}{2\omega_{1}(t)}-\frac{\dot{\gamma}(t)}{2\omega_{2}(t)}~.
\end{equation}
With this choice, we finally arrive at a concise expression of Berry's phase:
\begin{equation}
    \phi^{(n_{1},n_{2})}_{B}=(n_{1}+\frac{1}{2})\phi^{(1)}_{g}+(n_{2}+\frac{1}{2})\phi^{(2)}_{g}~.
    \label{k}
\end{equation}
Furthermore, the above phase factor is of purely quantum origin and is a  phase over and above the dynamical phase
and the classical counter part of the Berry phase \cite{H}  simply read off as
\begin{equation}
  \phi_{H}= -[\frac{\partial}{\partial n_{1}}  +\frac{\partial}{\partial n_{2}}]\phi^{(n_{1},n_{2})}_{B}=-(\phi^{1}_{g}+\phi^{2}_{g})~,
\end{equation}
which was originally established by Berry \cite{berry1}. As a result, it is worth noting that our geometric phase factor in the Schroedinger picture (\ref{bg}) coincides exactly to the additional phase shift established beyond the dynamic phase while adiabatically transporting  the ladder operators of our system Hamiltonian.


\section{Estimation of BP with respect to frequency of oscillator}\label{AppE}
\renewcommand{\theequation}{E.\arabic{equation}}

We choose two slightly an-isotropic time dependent frequencies to demonstrate the exact expression of the low frequency gravitational wave induced Berry phase and its variations with detector frequency amplitude. For that we take the following structures for $\Omega_{1}(t)$ and $\Omega_{2}(t)$:
\begin{equation}\label{choicef}
\Omega^{2}_{1}(t)=(\omega_{0}+\Omega_{0}cos(\omega_{g}t))^{2}+\nu^{2}_{0}sin^{2}(\omega_{g}t);~~~\Omega^{2}_{2}(t)=(1+\delta)\Omega^{2}_{1}-\delta \nu^{2}_{0} sin^{2}(\omega_{g}t)
\end{equation}
with the small anisotropic parameter $\mid\delta\mid<<1$ and $\omega_{0}>\Omega_{0}$, we maintain the positivity of $\omega_{1}(t)$ and $\omega_{2}(t)$ for all time $t$. Notably, $\omega_{g}$ characterizes the angular frequency of low-frequency gravitational waves, represented as $\omega_g=2\pi \nu_g$. Furthermore, we regard $\Omega_{0},\omega_{0}$ and $\nu_{0}$ as time-independent and adjustable parameters. Additionally, we introduce the gravitational wave interaction coupling parameter, denoted as $\gamma=\omega_{g}\chi_{0}\tilde{\epsilon}_{+}sin(\omega_{g}t)$.
 The above choices guarantee the synchronization of the time periods of the detector's frequency parameters with the low-frequency gravitational wave's frequency. Such a choice is very important which we discussed below Eq. (\ref{PRD1}).

Furthermore, it should be highlighted that the Berry phases $\phi^{(1)}_{g}$ and $\phi^{(2)}_{g}$ (given in Eqs. (\ref{finalres}) and (\ref{finalres2})), will be of the same order of magnitude because of the small anisotropy in the instantaneous frequencies that correspond to each mode for each of the respective arms. 
Practically, while undergoing adiabatic transport along a closed circuit in parameter space, it's crucial for the integrand of the phase factors, along with the dynamical phases described in (\ref{finalres}) and (\ref{finalres2}), to remain finite and real at each moment throughout this time period to ensure the validity of the adiabatic theorem. For the convenience of computation of the integral (\ref{finalres}) and (\ref{finalres2})  can be first rewritten as
\begin{equation}
    \phi^{(1)}_{g}=\int^{t=\frac{2\pi }{\omega_{g}}}_{t=0} \frac{\omega^{2}_{g}\tilde{\epsilon}_{+}\chi_{0}  cos(\omega_{g}t)}{\omega_{0}+\Omega_{0}cos(\omega_{g}t)} dt;~~~  \phi^{(2)}_{g}=-\frac{1}{\sqrt{1+\delta}}\phi^{(1)}_{g},
\end{equation}
and, then (\ref{finalres})  can be again recasted in-terms of uni-modular complex parameter $z(t)=e^{i\omega_{g}t}$ as
\begin{equation}
    \phi^{(1)}_{g}=\frac{-i\omega_{g}\chi_{0}\tilde{\epsilon}_{+}}{\Omega_{0}}\oint_{\mid z\mid=1} dz~ \frac{(z^{2}+1)}{z(z^{2}+2az+1)};~~~~~~a=\frac{\omega_{0}}{\Omega_{0}},
    \label{F}
\end{equation}
where we see that the above integral reduces to a simple loop integral over the unit circle in the complex plane. Note that here we have taken $\nu_{0}=2\omega_{g}\chi_{0}\tilde{\epsilon}_{+}$ for simplicity.

In this context, we would like to mention that this alternative complex re-parameterization of our ``parent" time-dependent parameter space spanned by $\gamma$ and $\Omega_{i}$ (occurring in $\phi^{(i)}_{g},~ i=1,2$) has enabled us to obtain this above  simplified form. In fact the following  simple identity
\begin{equation}
\frac{(z^{2}+1)}{z(z^{2}+2az+1)}=\frac{1}{z}+\frac{a}{\sqrt{a^{2}-1}} \big(\frac{1}{z-z_{-}}-\frac{1}{z-z_{+}}\big)~,
\end{equation}
helps us to identify the three simple poles  in the integrand (\ref{F}) as
\begin{equation}
  z=0, ~ z= z_{\pm}= -a\pm \sqrt{a^{2}-1}~.
\end{equation}
Out of which only $z=0$ and $z=z_{+}$ lie within the unit circle whereas $z=z_{-}$ lies outside for $a>1$. This follows trivially from the fact  that $z_{+}|_{a=1}=-1$, and $\frac{dz_{+}}{da}=-1+\frac{a}{\sqrt{a^{2}-1}}>0$,  $\forall a > 1$. We can therefore disregard $z_{-}$ completely to compute the above integral $(\ref{F})$ in a straightforward manner to obtain the  Berry phase
\begin{equation}
\phi^{(1)}_{g}=\frac{2\pi\omega_{g}\chi_{0}\tilde{\epsilon}_{+}}{\Omega_{0}}[1-\frac{1}{\sqrt{1-\epsilon}}]~.
\label{F5}
\end{equation}
Here we have introduced $\epsilon=\frac{1}{a^{2}}$ fulfilling the condition  $0<\epsilon<1$.

\color{black}
\vspace{0.9cm}
\begin{figure}
\includegraphics[scale=0.80]{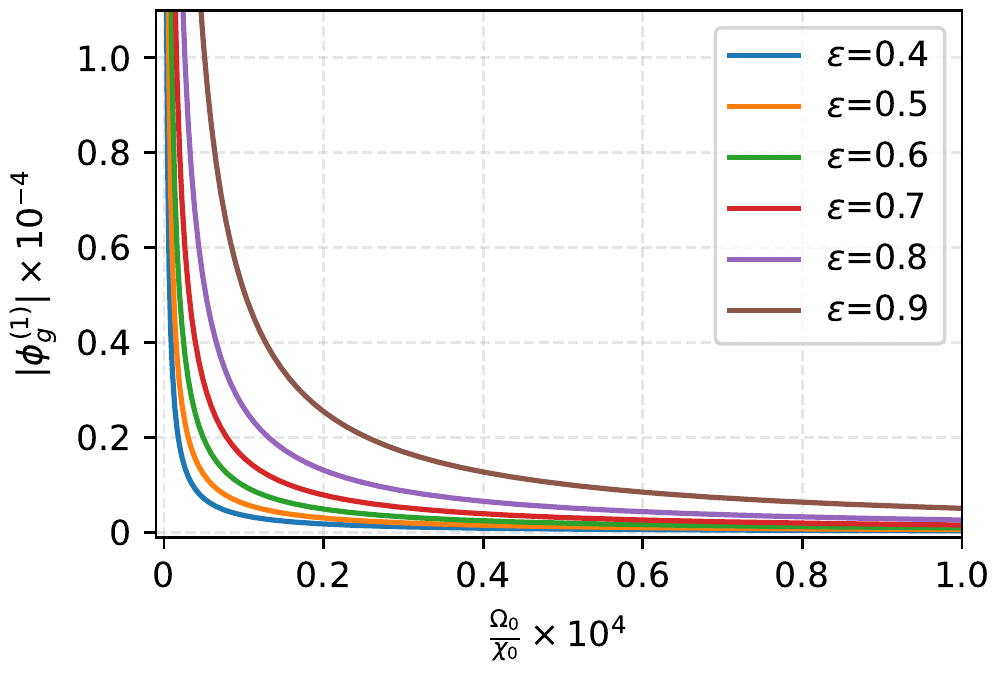}
\caption{A schematic representation showing the magnitude of GWs-induced Berry phase vs the detector's scaled frequency range with LFGWs amplitude $\chi_0=10^{-21}$ and frequency $\nu_g=0.01Hz$, polarization $\tilde{\epsilon}_{+}=1$}
\label{Fig1}
\end{figure}

In this context we can mention that this integral (\ref{F}) can also be computed alternatively by using the poles of the integrand at $z_{-}$ and $z=\infty$ (equivalently at $w=0$ for $w=\frac{1}{z}$) which are also enclosed by the above unit circle if the function is represented on the compactified  Riemann sphere-albeit in the opposite orientation.
It is important to highlight that although the system can never acquire the specific parameter values ($z_{0}$ and $z_{\pm}$), these values can still have an impact on the contour integral due to the nonholomorphic nature of the integrand (\ref{F}) at these simple poles within the contour ($\mid z\mid =1$).

Finally, it may be noted that all the  closed contours $C$ in the above-mentioned parent parameter space, associated with the parameters ($\Omega_{1},\gamma$), can take different sizes/shapes depending upon the free parameters $\Omega_{0},\omega_{0}$ and also on $\omega_{g}$. Interestingly, however, all such closed contours get mapped to the same unit circle: $\mid z\mid=1$ in the complex $z$-plane.  With this the phase integral $(\ref{F})$ gets determined \textit{almost} uniquely up to an overall constant determined by the ratio of the angular frequency of the external gravitational wave ($\omega_{g}$) and that of the constant parameter $\Omega_{0}$  occurring in ($\ref{F5}$): $\frac{\omega_{g}}{\Omega_{0}}$. Any deformation in contour $C$ will result in  shifting the poles $z_{+}$ (with $a>1$) in the unit circle ($\mid z\mid=1)$ and will change the value of phase integral (\ref{F5}).
Moreover, the graphical representation of the BP, shown in Fig. \ref{Fig1}, suggests that the nonzero finite magnitude of Berry phase is induced by GWs, which is crucial given that the detector frequency range is in the ultra-low  ($\Omega_{0}\sim10^{-17}rad/sec$) frequency range \cite{UltraF}.







\end{appendix}
\end{widetext}

\end{document}